\documentclass[journal]{IEEEtran}

\usepackage[utf8]{inputenc}
\usepackage[T1]{fontenc}
\usepackage{graphicx}
\usepackage{url}
\usepackage{siunitx}
\usepackage{xcolor}
\begin{document}

\title{A 9T4R RRAM-Based ACAM for Analogue Template Matching at the Edge}

\author{Georgios Papandroulidakis,
        Shady Agwa,~\IEEEmembership{Member, IEEE},
        Ahmet Cirakoglu,~\IEEEmembership{Member, IEEE}
        \\and~Themis Prodromakis,~\IEEEmembership{Senior Member, IEEE}\\
        Centre for Electronics Frontiers (CEF), Institute for Integrated Micro and Nano Systems\\University of Edinburgh, UK
\thanks{G. Papandroulidakis, S. Agwa, A. Cirakoglu and T. Prodromakis are with the Centre for Electronics Frontiers (CEF), Institute for Integrated Micro and Nano Systems, University of Edinburgh, UK \\
E-mails: \{gpapandr, shady.agwa, t.prodromakis\}@ed.ac.uk, a.cirakoglu@sms.ed.ac.uk\\
Contact e-mail: gpapandr@ed.ac.uk}
\thanks{The authors acknowledge the support of the EPSRC FORTE Programme (Grant No. EP/R024642/2), the EPSRC ProSensing Project (Grant No. EP/Y030176/1) and the RAEng Chair in Emerging Technologies (Grant No. CiET1819/2/93).}
\thanks{}}

\IEEEoverridecommandlockouts
\IEEEpubid{\begin{minipage}[t]{\textwidth}\ \\[8pt]
        \centering\normalsize{A patent application in relation to this technology has been submitted. \\ This work has been submitted to the IEEE for possible publication. \\ Copyright may be transferred without notice, after which this version may no longer be accessible.}
\end{minipage}}

\markboth{}%
{G. Papandroulidakis \MakeLowercase{\textit{et al.}}: A 9T4R RRAM-Based ACAM for Analogue Template Matching at the Edge}



\maketitle

\begin{abstract}
The continuous shift of computational bottlenecks to the memory access and data transfer, especially for AI applications, poses the urgent needs of re-engineering the computer architecture fundamentals. Many edge computing applications, like wearable and implantable medical devices, introduce increasingly more challenges to conventional computing systems due to the strict requirements of area and power at the edge. Emerging technologies, like Resistive RAM (RRAM), have shown a promising momentum in developing neuro-inspired analogue computing paradigms capable of achieving high classification capabilities alongside high energy efficiency. In this work, we present a novel RRAM-based Analogue Content Addressable Memory (ACAM) for on-line analogue template matching applications. This ACAM-based template matching architecture aims to achieve energy-efficient classification where low energy is of utmost importance. We are showcasing a highly tuneable novel RRAM-based ACAM pixel implemented using a commercial 180 nm CMOS technology and in-house RRAM technology and exhibiting low energy dissipation of approximately 0.036 pJ and 0.16 pJ for mismatch and match, respectively, at 66 MHz with 3.3 V voltage supply. A proof-of-concept system-level implementation based on this novel pixel design is also implemented in 180 nm.
\end{abstract}


\IEEEpeerreviewmaketitle

\section{Introduction} \label{sec:introduction}
\IEEEPARstart{I}{N} recent years, the main driving technology of electronic circuits and systems, the MOSFET technology, is showcasing important difficulties in keeping up with the increasingly higher computing demands through scaling \cite{Hamdioui2016}. Simultaneously,it has been shown that one of the most inhibiting factors in modern systems' performance is the data movement that occurs in modern highly distributed edge computing networks, where a large amount of interconnected sensors, edge computers and servers are constantly communicating transferring huge amounts of data \cite{Wilkes1995, McKee2004, Zidan2018}. \par
Recently, a lot of focus has been given to novel wearable and implantable edge devices aimed at sensing specific bio-signals and transferring information about health and activity \cite{Wang2008, Eggimann2021, Menon2021}. Brain Machine Interface (BMI) devices are one of these novel edge computing applications for monitoring brain activity \cite{RAPEAUX2021102}, a process that can be helpful in advanced medicine, such as better understanding and treating brain-based diseases as well advancing the frontier of neural prosthetic development \cite{neuroprosthesis_ref_nature}. A cornerstone operation that enables BMI implementations is real-time template matching, where an analogue signal can be quickly compared against a set of predefined templates and potentially identified as one of them. This enables a fast and low power classification operation near the signal's origin. \par
\begin{figure*}[t!]
	\centering
	\includegraphics[width=15.8cm]{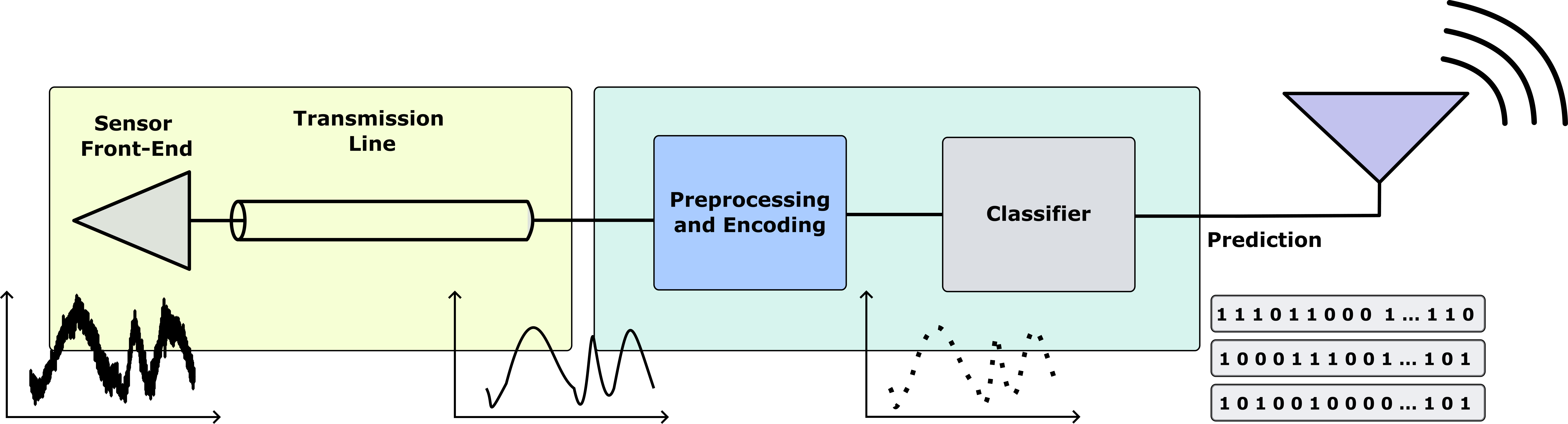}
	\caption{Concept-level diagram of the TXL ACAM application in near-sensor implementations for signal classification. The signal online template matching application envisioned for the proposed RRAM-CMOS ACAM assumes the signal capturing and pre-processing through appropriate analogue front-end circuitry and then encoding of the signal into an appropriately formatted input query to search through the RRAM-based ACAM classification engine back-end. An analogue input vector sampled from the singular continuous signal is supplied to the classifier as a query key.}
	\label{fig:ACAM_concept}
\end{figure*}
Template matching approaches enables the implementation of novel low-power pattern recognition engines capable of being closely integrated with edge sensors. Content Addressable Memory (CAM) is the main system employed for many template matching applications \cite{Ni2019, Alkabani2019, Eshraghian2011, Yin2020d, Graves2020}. CAMs are memory-centric systems that are deploying large memory arrays with hardware-level optimisation that enable fast parallel search and match capabilities. Their template match operations are based on a query input that is compared (in a massively parallel manner) with all stored templates of the CAM array \cite{Graves2019}. The general architecture of CAMs is based on the implementation of memory arrays with each cell of the array being modified to perform an in-situ comparison between an input signal and the contents of the memory cell with each row being organised as a template \cite{Yang2020b, Tanaka2020}. By providing an input query vector, we are able to perform a parallel comparison of this input with every template. The sensing front-end circuitry of an edge application usually deals with analogue stimulus from its environment, therefore consuming high energy and latency penalty for analogue-to-digital conversion (ADC). Analogue CAMs (ACAMs) \cite{Graves2019} can play an important role in eliminating the analogue/digital cross-interfacing penalty, where analogue inputs can be classified on-the-fly without converting them into the digital domain. This reduces not only the design complexity but also the data-transfer bandwidth, enabling the implementation of novel energy efficient always-on classification engines at the edge.  \par
At the same time, advances in emerging memory technologies introduce a new component for novel analogue circuit design. Memristors, also known as Resistive Random Access Memory (RRAM), are two-terminal, tuneable, non-volatile, nanoscale resistive memory devices \cite{Chua2014, Ielmini2018}. RRAM has many beneficial traits over conventional memory devices \cite{Sebastian2018, Sebastian2020, Stathopoulos2017a}. RRAM can be integrated to implement novel computing systems showcasing low power consumption, high throughput, low area of integration and multi-bit information stored per cell compared to their fully CMOS counterparts \cite{Sebastian2020, Zhang2020}. Thus, RRAM is considered as an excellent candidate for building ACAMs. Each RRAM device is capable of storing multi-bit information per single device in the form of resistance \cite{Graves2018}. The different resistance values form the analogue matching windows of each ACAM cell \cite{Serb2018}. \par
Fig. \ref{fig:ACAM_concept} shows the concept of the proposed system where the ACAM-based template matching approach is used to perform classification on the edge, tightly coupled with the sensing front-end. In this case, the ACAM implements the classifier as shown in Fig. \ref{fig:ACAM_concept}. The dataflow of showcased paradigm includes capturing analogue signals through a specialised sensor, preprocessing this stimulus and then providing it in form of queries to the template matching classifier. The front-end circuitry detects and amplifies the bio-signals and then forwards them to the near-sensor RRAM-based ACAM edge classifier. The detected signals will be sampled and compared to the pre-trained templates to reach a classification decision. Based on the templates stored in ACAM, a match window per cell can be defined with the query value connected to each cell being compared with this memory-defined window. Different matching conditions can be implemented when assessing the results of ACAM through exact match, best match and threshold match. For different matching instructions a different matching comparator network needs to be implemented (e.g., a Winner-Take-All Network for best match) Depending on the applied matching policy followed by the implementation, different outputs can result from the classifier. In general, the more inputs of the query vector falls within the matching windows of the template's cells, the higher the matching level of the query with the specific template. \par 
In this work, we are showcasing a novel RRAM-based analogue Template matching piXeL (TXL) used to implement an energy efficient ACAM system (TXL-ACAM) towards accelerating analogue near-sensor template matching operations at the edge. We are showcasing a hybrid CMOS-RRAM IC using 180nm CMOS technology and in-house RRAM technology integrated as back-end-of-line (BEOL). The implemented array of the prototype TXL-ACAM is $32\times48$ array. The  TXL-ACAM integrates a $32\times32$ array, with 2K RRAM devices, and a $16\times32$ polysilicon-based emulator array. The area of integration for the prototype  TXL-ACAM is approximately $3.8 mm^{2}$. The 9T4R TXL design is using 2 RRAM reconfigurable devices while the other 2 resistive elements are implemented as conventional polysilicon devices. The cell design can exhibit competitive energy dissipation of approximately $170 fJ$ in worst case scenario (matching cell for input close to the operating voltage of 3V). Peripheral circuit is also implemented enabling the programming and readout of the RRAM-based ACAM array. The proposed TXL-ACAM energy efficient inference engines can reduce the data transfer between sensors and higher performance systems by classifying information at the edge. \par 
The rest of the paper is organised as follows: In Section \ref{sec:background}, we present the main background of the RRAM-based analogue and digital CAMs. In Section \ref{sec:9T4Rcircuit}, we showcase the proposed TXL ACAM cell design and operation, and its performance as well as comparison with state-of-art. In Section \ref{sec:prototypeIC}, we exhibit the hardware implementation of a TXL-based ACAM array alongside the necessary peripherals. Finally, in Section \ref{sec:conclusions}, we discuss our findings and conclusions.  \par
\section{Background} \label{sec:background}
In the last decade, many different RRAM-based circuits and systems have been proposed that introduce information processing at the analogue domain. RRAM is mainly aimed at memory-centric applications where RRAM's key capabilities, such as storing multiple bits of information per devices and being non-volatile, enable the development of novel electronics that can better emulate neuro-inspired operations. One such neuro-inspired in-memory computing application is Associative Memory and its main digital emulation the CAMs \cite{Kohonen1980}. Although digital or ternary CAMs have seen a lot of development, analogue CAMs' (ACAMs) development, which can be used as a better analogue to the associative memory recall systems observed in biological neural networks, was inhibited by the lack of a appropriate analogue nanoscale device. The integration of RRAM devices into CAMs has re-invigorated the design of neuro-inspired associative recall functions through the development of novel ACAMs \cite{Pedretti2021}. ACAMs can be leveraged to perform complex look-up operations through large sets of stored templates. Additionally, they can calculate the similarity of an input with each one of these templates through approximate computing techniques in the analogue domain \cite{Pedretti2021,Pedretti2022}. The use of RRAM-based circuits capable of storing and processing multi-bit information enabled the close integration of such systems with the sensors and pre-processing analogue circuits, that capture the stimulus from the environment, since no conversion to binary is required. Thus, processing the information without the need to transfer them into the digital domain is possible, minimising a process known for being costly in terms of IC area, power and delay. \par
There is a variety of different pixel designs for implementing hybrid RRAM-CMOS CAM systems. Compared to the conventional fully CMOS CAM cells, usually configured as SRAM-based circuits \cite{Ahmed2017}, RRAM-CMOS cells showcase better power efficiency, lower area of integration, non-volatility and capability in storing multi-bit information \cite{Junsangsri2017, Graves2020, Graves2019}. Towards providing novel solutions for enhancing CAM and ACAM systems, many different hybrid CAM cell designs using emerging memory technologies have been proposed in recent years. In \cite{Li2020_HP_ACAM}, a 6T2R ACAM cell is presented that uses two 1T1R voltage divider circuits to map the lower and upper thresholds (that determines the matching window of the cell). Additionally, another four transistors implementing a CMOS inverter to enable complementarity of the upper threshold control in defining the matching window of the cell and two nMOS transistors for the per cell comparison are also used for a total of 6T2R design. The design exhibits good performance and tuneability with each two bounds mapped as resistance states in the two RRAM devices used. ACAM cells such as the one exhibited in \cite{Li2020_HP_ACAM} use resistive voltage dividers (through 1T1R) to read the memory and perform the per cell match operation. The use of 1T1R in general results in smaller pixel size and thus increases array density but introduces static power consumption during ACAM search operation, since it requires to apply a reference readout voltage to the two 1T1R branches. The per cell comparison uses the intermediate voltages of the two 1T1Rs which are creating a direct resistive path between the reference readout voltage and ground. Techniques to minimise the static power dissipation by applying a small reference readout voltage or by using large resistive states for RRAM could be applied but might not be possible for different RRAM technologies. Additionally, by using CMOS design to map the cell threshold, instead of the conventional 1T1R branches, we could eliminate the static power during the search operation while retaining the capability of programming arbitrarily matching windows per cell. Saving static power through hybrid RRAM-CMOS logic could provide an interesting trade-off between power and pixel size that depending on the application could be worth considering. \par
In \cite{Agwa2023_3T1R_ACAM}, a 1T1R+2T pixel design for ACAM is showcased as a novel form of less complex ACAM cell requiring half the area compared to the current state-of-art 6T2R ACAM cell design \cite{Graves2020}. The 1T1R+2T cell is one of the compact designs for ACAM systems comprised of only 3 transistor devices and one RRAM device acting as the memory element. Although the area and energy of the 1T1R+2T shows promising results, it has some limitations in terms of tuneability of its low and high threshold. This is due to the design using only one RRAM device and 2 MOSFET devices to effectively trigger the thresholds based on the voltage threshold of the MOSFET technology employed. In \cite{Serb2018}, a 6T2R template matching cell is proposed for event generation when a specific voltage value is provided as stimulus. This design showcases good power consumption during template matching operation and exhibits competitive area usage but it has some limitations in terms of configurability of the matching window as well as with accessing the RRAM devices for programming, thus requiring some additional access circuit. Additionally, the cell's response is dependent on a current mirror that increases the amplitude of the generated spike (used to logically map the event) which will require careful calibration and design to avoid having substantially different spike generation amplitudes in a large array. \par 
\section{9T4R Cell Design and Operation} \label{sec:9T4Rcircuit}
The TXL circuit has the topology of two hybrid RRAM-CMOS 2T2R inverters and an additional CMOS-based output network controlled by the two 2T2R hybrid inverters. The input signal is connected to the input of the RRAM-CMOS inverters (gate nodes of the MOS components in a conventional CMOS inverter connectivity). The RRAM devices are placed between the pMOS and $V_{DD}$ connection and between the nMOS and the GND node for the pull-up and pull-down network of the hybrid inverter, respectively. The output of one hybrid inverters controls the gate of the output pMOS device and the output of the other hybrid inverter controls the gate of the output nMOS device. Through this configuration, a comparison of the input with the stored matching window (determined by RRAM conductance states) is possible. \par
\begin{figure*}[t!]
	\centering
	\includegraphics[width=17cm]{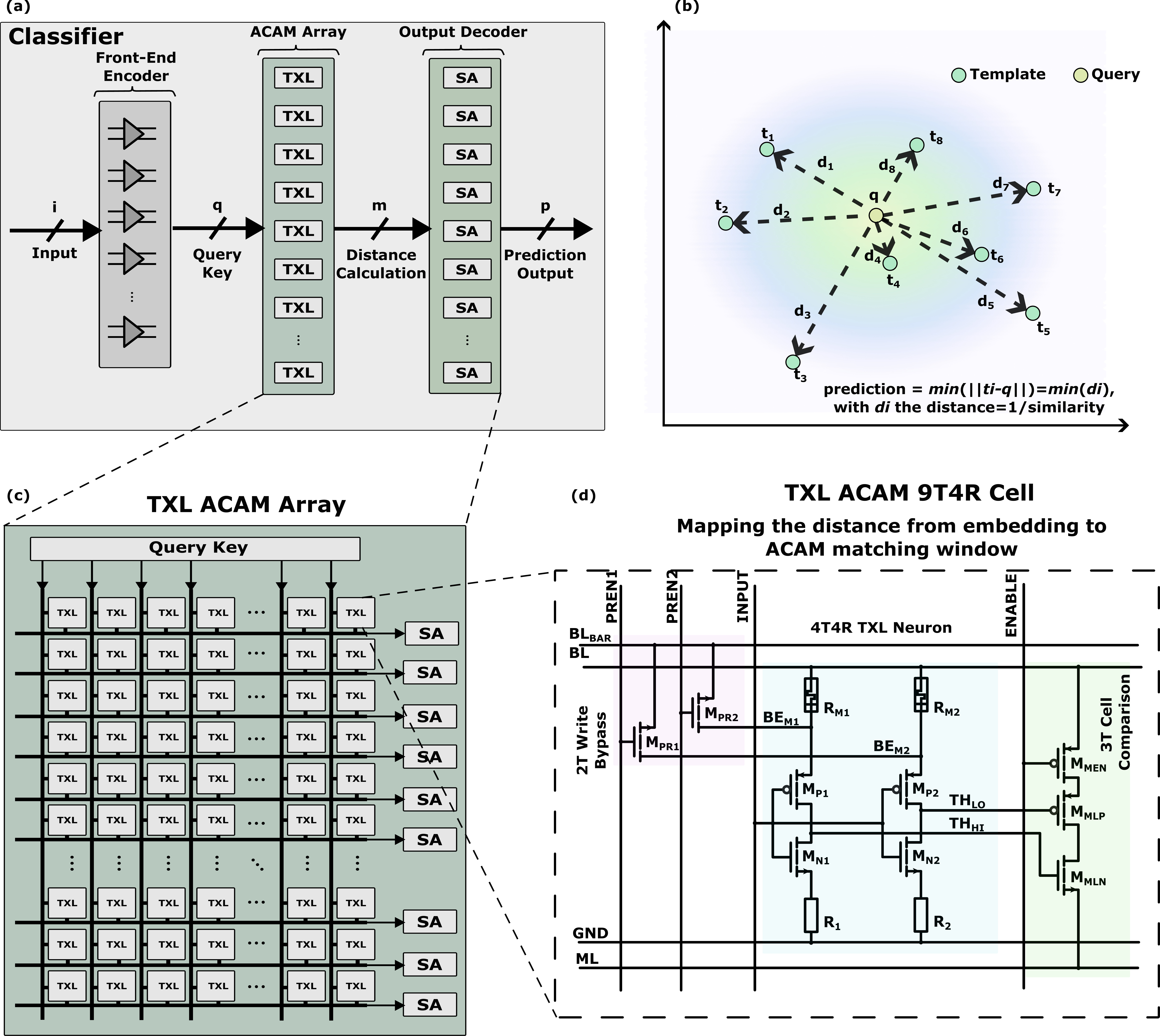}
	\caption{(a) General diagrams of the operation layers used for ACAM (b) ACAMs are used to calculate the distance between the query input and the patterns stored in the ACAM. (c) General TXL array organisation with the query being distributed to the columns of the array to enable a parallel search operation while all rows of the array are organised as matchlines that calculate and send to the sense amplifiers (SAs) the distance of the query and the template stored in this matchline. (d) Circuit schematic of the proposed 9-transistors-4-resistors (9T4R) pixel design for analogue template matching applications. The cell can map two threshold (a low and a high threshold) through the use of its non-volatile RRAM devices ($R_{M1}$ and $R_{M2}$). The use of the hybrid RRAM-CMOS inverter design enables the movement of the threshold voltage of the inverter depending on the resistive ratio of $R_{M1}$-$R_{1}$ and $R_{M2}$-$R_{2}$ (through source degeneration of the MOSFET devices). The memory part of the cell is comprised from the two RRAM-CMOS inverters while the per cell readout circuit (3T Cell Comparison part) is comprised from $M_{MEN}$,$M_{MLP}$ and $M_{MLN}$ devices. The $M_{PR1}$ and $M_{PR2}$ devices of the 2T write bypass circuit are used for accessing the RRAM for programming.}
	\label{fig:9T4R_Cell_Schematic}
\end{figure*}
\begin{figure*}[t!]
	\centering
	\includegraphics[width=16.5cm]{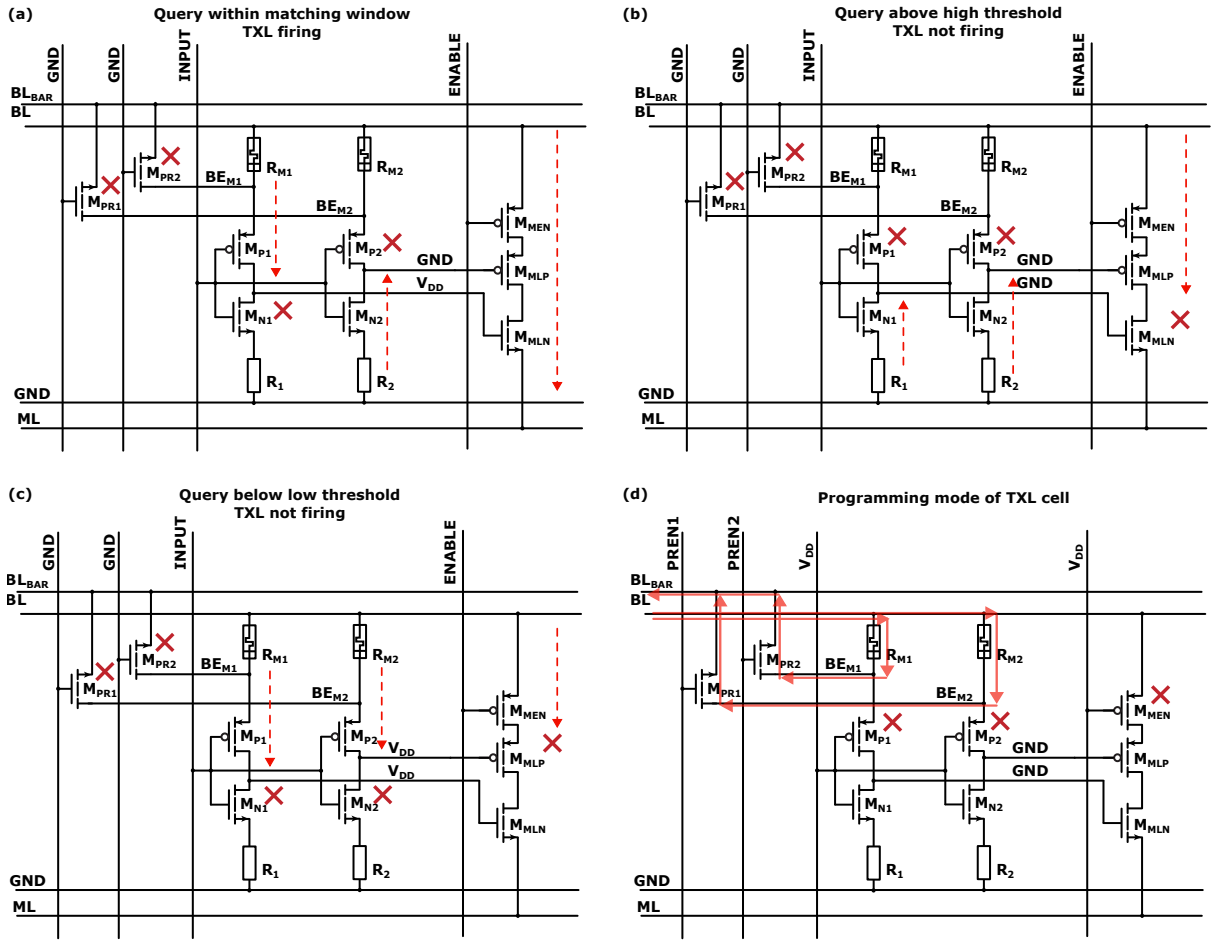}
	\caption{TXL modes of operation. In (a), (b) and (c) three different match/mismatch cases are shown while in (d) the programming mode has been enabled with the 2T2R and 3T branches no used and the 1T1R programming paths enabled. The 9T4R cell operates effectively in two modes: template matching mode and programming mode. During the template matching mode the $M_{PR1}$ and $M_{PR2}$ nMOS are non-conductive and the cell is using its RRAM conductance states to compare an input with its internal matching window. While during programming mode, the  $M_{PR1}$ and $M_{PR2}$ nMOS are conductive (one at a time) to enable a low resistance path to close the appropriate circuit with its peripherals and write the RRAM devices to specific conductance states.}
	\label{fig:9T4R_Cell_Modes_of_Operation}
\end{figure*}
\begin{figure}[h]
\includegraphics[width=7cm]{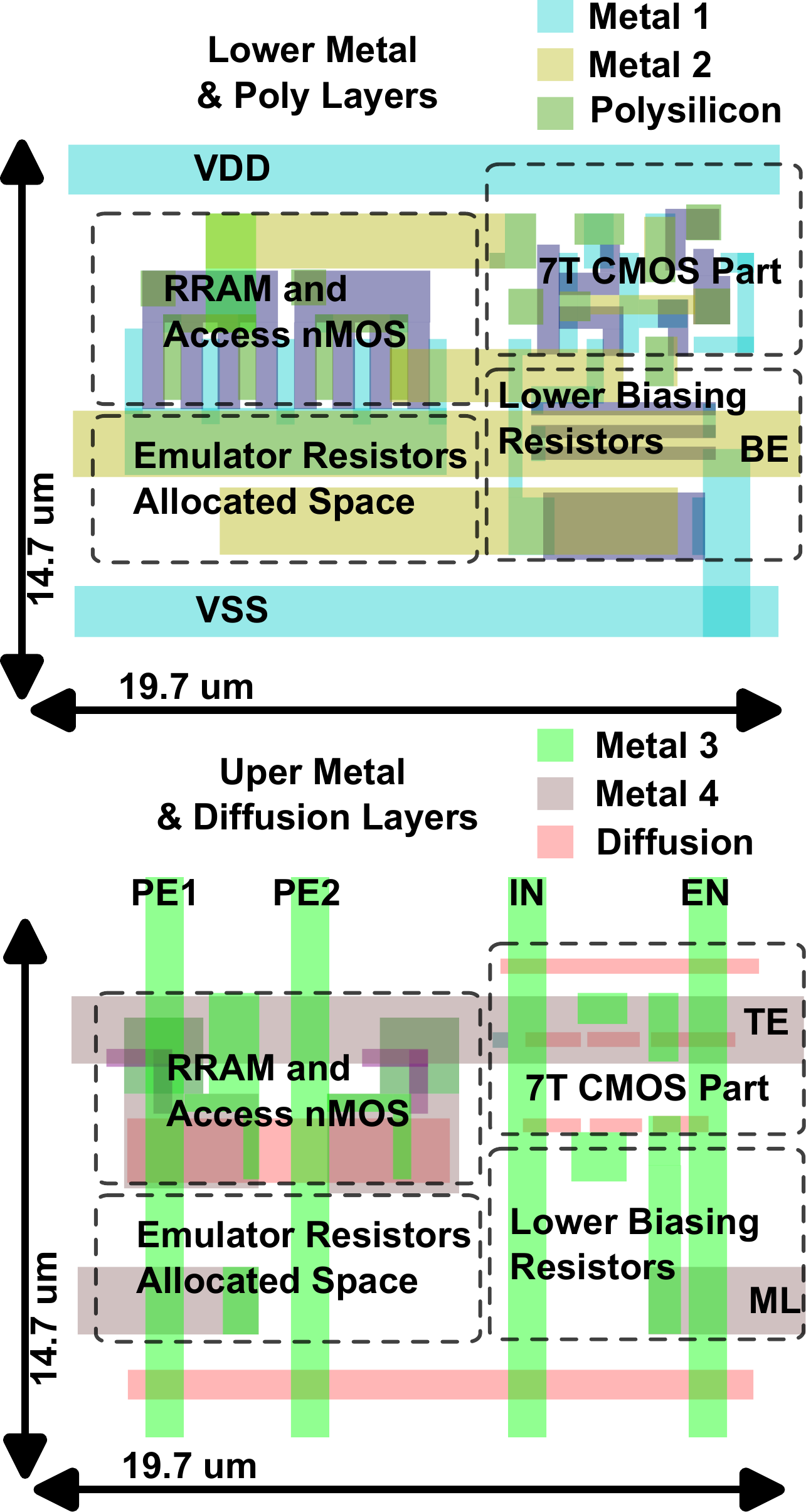}
\centering
\caption{In this figure, a representation of the 9T4R cell layout is shown alongside its dimensions. the main parts of the cell consist of the 7T CMOS part of the cell and the 2T programming nMOS devices alongside the BEOL integrated RRAM devices. The lower left area of the cell is allocated for polysilicon integration for the emulator array where instead of BEOL RRAM integration, conventional polysilicon fixed resistors are used for testing and calibration. The lower right area is used for the biasing resistors $R_{1}$ and $R_{2}$.}
\label{fig:9T4R_layout_design}
\end{figure}

In Fig. \ref{fig:9T4R_Cell_Schematic}(d), a schematic representation of the 9T4R ACAM pixel is shown. The memory part is based on the use of two hybrid RRAM-CMOS inverters (2T2R circuits) with each such hybrid inverter mapping one of the two thresholds/bounds. The threshold is effectively defined based on the ratio of the upper and lower RRAM device which shifts the voltage threshold of the hybrid inverter (thus the value of input voltage that the inverter changes output). The cell-wise comparison part is comprised of 3 MOSFET devices, one pMOS and one nMOS serially connected to multiply the low and high threshold memory read and one power-gating pMOS device that enables the charge of the matchline when the memory read operation is active and at the same time is performing a current limiting operation to appropriately adjust the charging rate of the matchline. This is performed by applying an appropriate biasing voltage ($V_{EN}$) to partially open the power-gating pMOS of the cell's comparator part. This per cell current limiter is useful for calibrating the cell charging capability depending on the size of the implemented array and/or the detection requirements of the accompanying sense amplifiers, thus it is an aspect of convenient configurability of the cell output. In Fig. \ref{fig:9T4R_layout_design} the layout design of the proposed 9T4R circuit is shown designed using 180 nm CMOS technology and a custom in-house RRAM parameterised cell (pcell) component \cite{Sachin_pcell}. It is worth noting that both RRAM-based and polysilicon-based 9T4R cells use the same general floorplan with the polysilicon-based emulator cells populating the lower left area of the cell as depicted in Fig. \ref{fig:9T4R_layout_design}. Thus, the dimensions and the floorplan of the MOSFET devices are the same for both RRAM-based and emulator cell designs. \par 
To enable an appropriate path for programming the RRAM devices $R_{M1}$ and $R_{M2}$, two additional MOSFET devices are used as access transistors during the programming/setup mode of the RRAM array where the RRAM devices are being set to the appropriate conductance states. If the cell is in programming mode, the path connecting the RRAM to the ACAM cell is being cutoff and the circuit uses these extra 2 nMOS devices to effectively operate the two RRAM as two 1T1R circuits. By accessing the RRAM devices between the $BL$ and $BE_{M1}$ and $BE_{M2}$ bottom electrodes provides a direct low resistance path for programming the RRAM devices. This negatively affects the density of the proposed ACAM cell design since it increases the necessary area of the cell, but the included programming specific accompanying nMOS devices enable proper access to program the RRAM devices. In this specific configuration, only the upper resistive elements are assumed to be programmable RRAM devices, while the lower resistive elements take the form of conventional polysilicon-based resistors. \par
\begin{figure}[t]
\includegraphics[width=\linewidth]{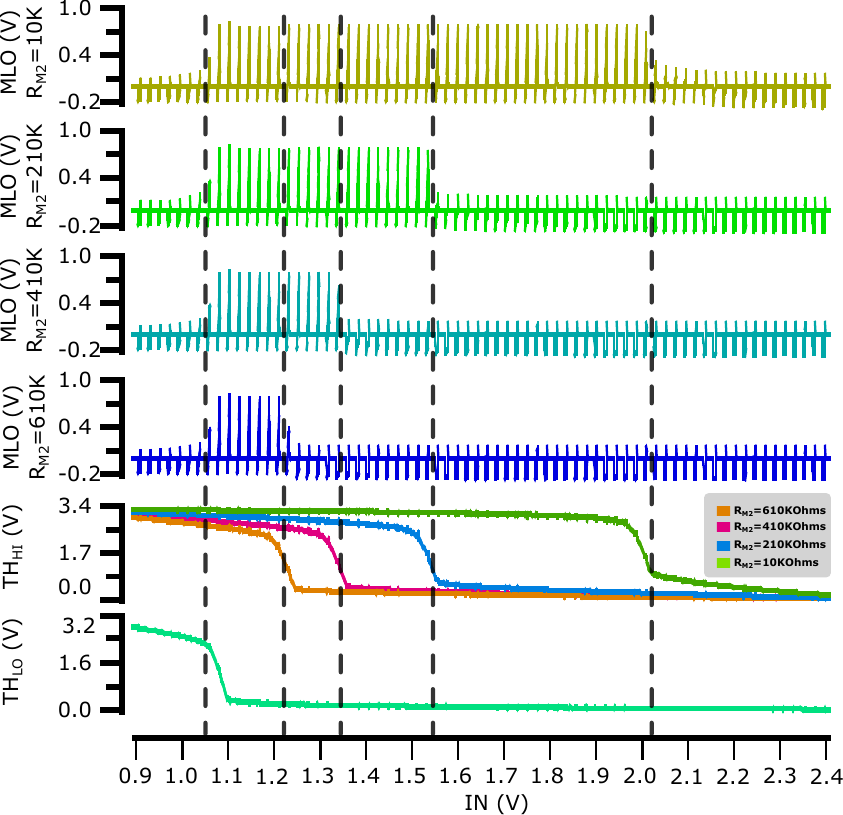}
\centering
\caption{Simulation results showcasing a typical matching window response for an input ramp. The analogue voltage input $V_{IN}$ is sweeped in transient simulation for different values of $R_{M2}$. The parametric analysis of the RRAM values are displayed on top of each other to showcase the moving of the upper threshold $TH_{HI}$. The simulations were performed in Spectre Cadence Virtuoso environment using 180 nm CMOS technology and in-house RRAM models \cite{Messaris2017}. It can be observed that for increasing $R_{M2}$ the high threshold is decreasing (parametric analysis of $TH_{HI}$) resulting in an increasingly smaller matching window (showcased by the output of the 9T4R through the $MLO$ set of traces), assuming that the lower threshold $TH_{LO}$ is not shifting due to changes in $R_{M1}$ value. The output of the cell for each RRAM configuration is shown in the $MLO$ set of traces. The charging and reset dynamics are controlled by the enable, reset matchline accumulator and operating voltage $V_{DD}$ =3.3 V.}
\label{fig:9T4R_matchin_window_response}
\end{figure}

In Fig. \ref{fig:9T4R_matchin_window_response} we can see the response of the proposed 9T4R circuit when sweeping the analogue voltage input for different RRAM configurations. By changing the configuration of the RRAM devices, thus programming them to different conductance states, different matching windows can be configured for each cell. The lower and upper threshold of these matching windows are controlled by the RRAM devices that change the resistive ratio of the 2T2R branches and can be seen in the lower two traces of Fig. \ref{fig:9T4R_matchin_window_response}. The upper threshold $TH_{HI}$ is shifting due to different $R_{M2}$ values while the lower threshold $TH_{LO}$ is the same since we are not changing $R_{M1}$ conductance. The enable signal is pulsed to perform the initialise-evaluation cycle of the circuit. The clock used for the simulation is $t_{enable}$ = 12 ns. For the transient analysis, a leakage resistor has been added to the output of 9T4R to reset the matchline charge. This leakage resistor is controlled by the $RES_EN$ signal which uses the same period of 12 ns to reset the output before the next evaluation. The voltage input is sweeped in the transient simulation while the RRAM values are sweeped through parametric analysis. We can see that for increased $R_{M2}$ values, we have a movement of $TH_{HI}$ to the left. \par
The circuit performs a pattern matching operation between its input signal and its stored patterns (stored in the form of programmable conductances through the integration of the non-volatile RRAM devices). Each hybrid inverter stores one of the bounds by appropriately configured RRAM devices. By convention, we consider that the actual bound is stored into one of the two RRAM devices (per hybrid inverter) while the other is being set to an appropriate mean value (between the lowest and highest conductance value attainable by the RRAM devices) and used as a reference conductance that enables an appropriate ratio in the hybrid voltage divider (hybrid inverter). The use of two RRAM devices per bound enable a better dynamic range for expressing conductance thresholds in our design. The reference RRAM devices are meant to be programmed once with a periodic refresh of their state to eliminate any conductance drift. The threshold RRAM devices are meant to be programmed each time a new set of patterns need to be mapped into the TXL-CAM system. \par 
After programming the RRAM-based conductances to appropriate values, the input signal (it could be either digital or analogue) is introduced to the input of the dual hybrid inverters mapping the two thresholds. The output of the hybrid inverters drives a custom pull-up network (consisting of pMOS and nMOS devices) which can detect the match of the input when sits between the lower and upper bound of the cell by connecting the high rail voltage (with $BL$ being connected to $V_{DD}$) to the matchline. In different case, we have a mismatch. The comparator part is power gated by a pMOS devices that is controlled by the global enable signal. The enable signal either cuts-off the power to the comparator part (during initialisation) or bias the power gating pMOS to a specific resistive state (during the evaluation phase). The biasing of the comparator branch effectively enforces a current limiting effect which results in a quick spike-like pulse of specific amplitude to appear in the matchline (for a given time of evaluation period). The charge limiting effect on the comparator branch has two effects. Firstly, it limits the power consumption per cell due to smaller voltage level transitions for the matchline. Secondly, it enables the analogue accumulation of charge in the matchline since we can define threshold above which the accumulated charge in the matchline triggers a match. The charge for each cell is accumulated using an analogue integrator circuit based on capacitor. If multiple cells have a match with the query input then multiple connection to $V_{DD}$ are enabled and the match line is driven at some specific rate (dependent on the number of match enable available per TXL-CAM array) to high voltage. The output sense amplifier reads the matchline and can detect if we have an (overall) match of the query to a specific TXL-CAM word. The sense amplifiers are calibrated to detect a specific voltage level which is translated to match based on the time-to-charge of the matchline. In case no such voltage level is observed within the readout operation timeframe, then the template is considered to not match with the input query. \par
\begin{figure}[t]
\includegraphics[width=7.8cm]{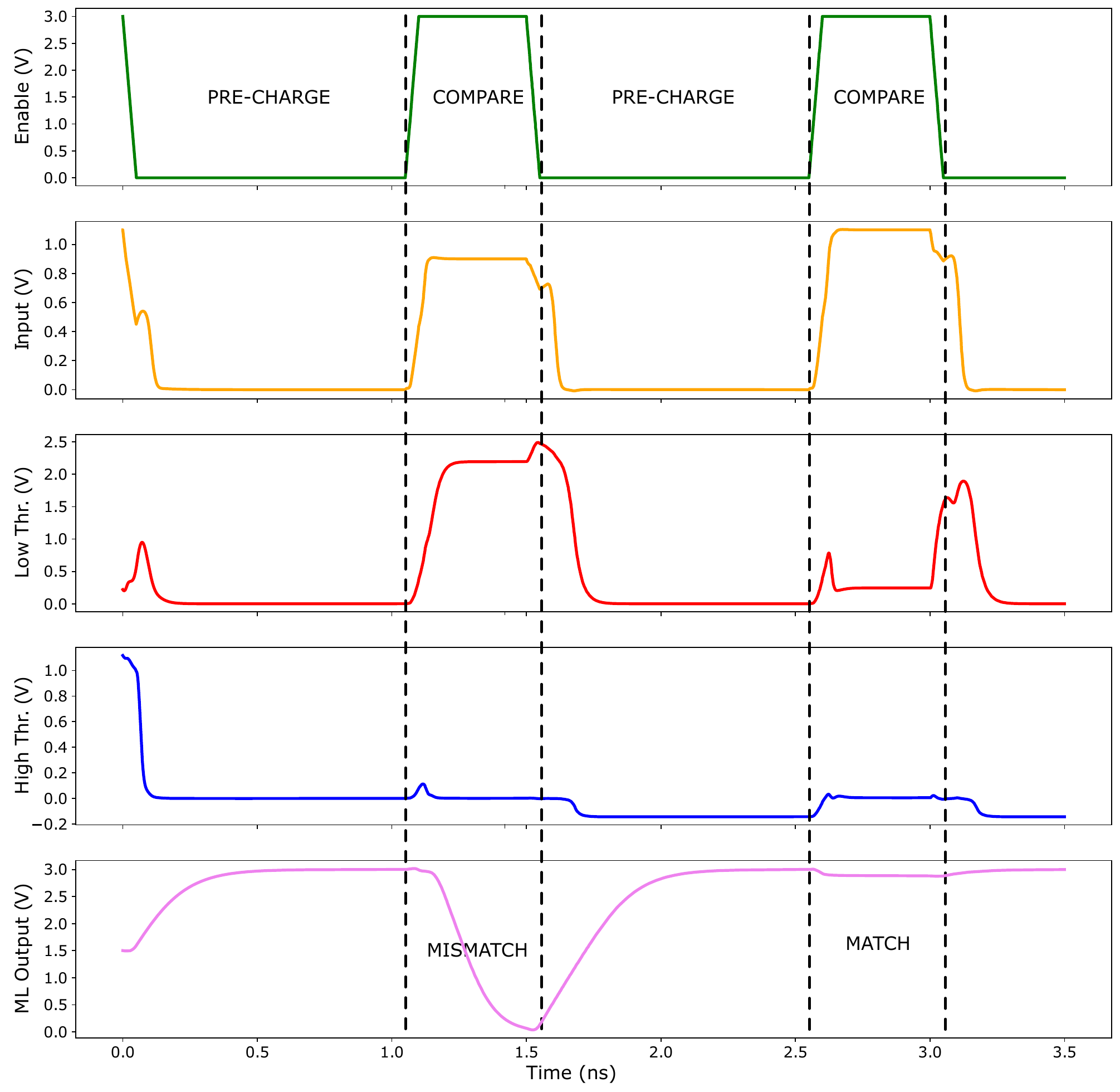}
\centering
\caption{Simulation results for a typical match and mismatch case for the state-of-art 6T2R which follows the conventional pre-charging matchline paradigm. More specifically, the figure illustrates the voltage of specific nodes in the ACAM cells. The simulations were performed using 5V components from a commercial 180nm CMOS technology. The 6T2R design from \cite{Li2020_HP_ACAM} was recreated in the abovementioned technology.}
\label{fig:6T2R_sim_res_voltage}
\end{figure}
\begin{figure}[t]
\includegraphics[width=7.8cm]{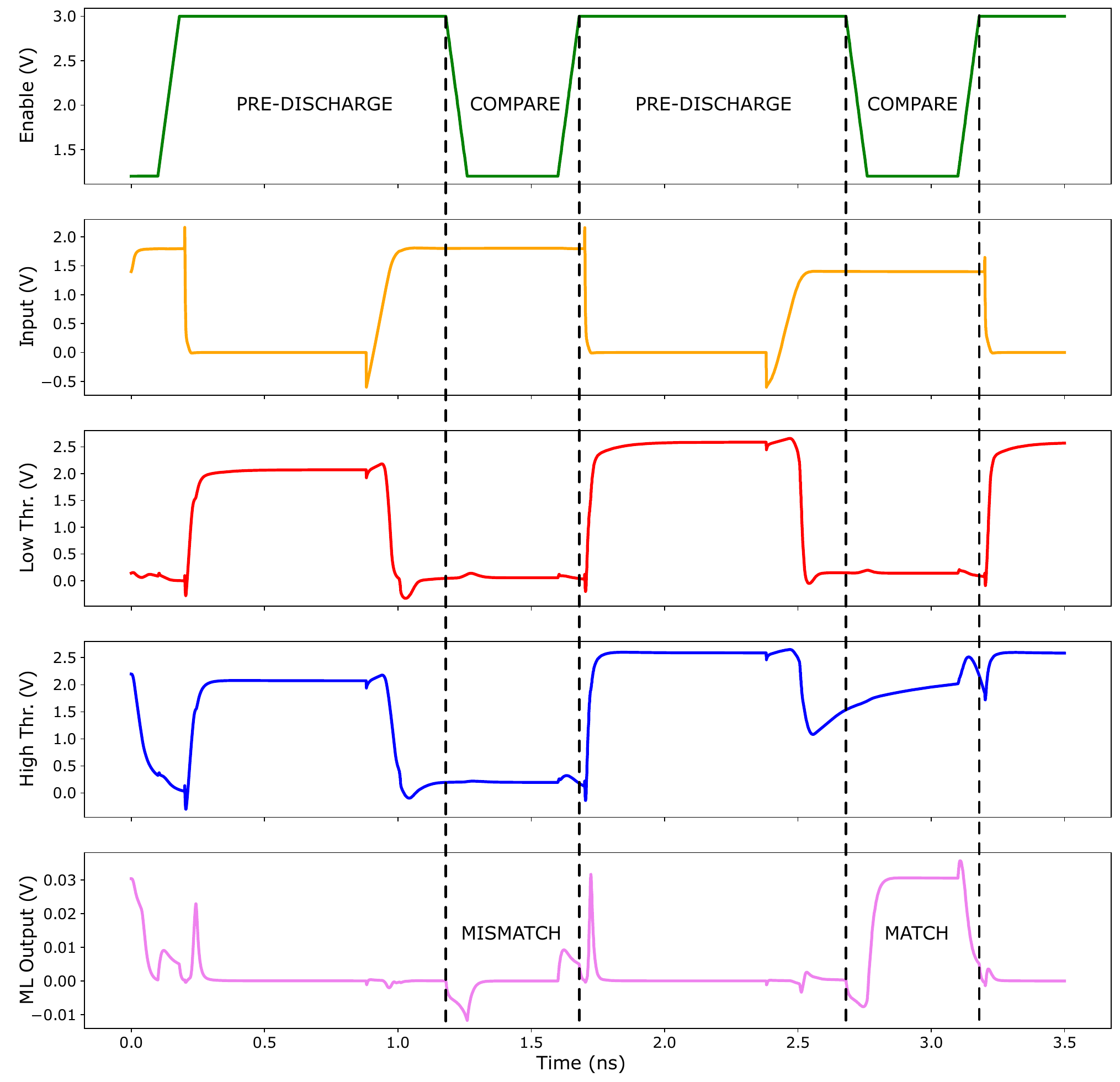}
\centering
\caption{Simulation results for a typical match and mismatch case for the proposed 9T4R which implements a conditional charging matchline paradigm, with the matchline being connected to GND before being charged to a specific voltage level based on the matching degree of the pattern matching operation. The discharged matchline that only charges upon memory hit can be considered advantageous energy-wise especially in applications with sparse hit events. More specifically, the figure illustrates the voltage of specific nodes in the ACAM cells. Although there is some noise in the matchline, the classification can be performed accurately by sampling within the pattern matching operation window, thus when the $TXL_{EN}$ is low. The simulations were performed using 5V components from a commercial 180nm CMOS technology. High voltage 5V MOSFETs were selected for the cell design to withstand the higher electroforming voltages that can be as high as $5V$.}
\label{fig:9T2R_sim_res_voltage}
\end{figure}
An important aspect of using a more complex design like the 9T4R against the state-of-the-art 6T2R is the use of hybrid RRAM-CMOS inverters as a method of reducing the dynamic power consumption of the template matching operation. The template matching operation in many RRAM-based cells usually involves the use of voltage dividers (i.e. through the use of 1-transistor-1-RRAM (1T1R) memory cells), thus creating a direct path between the high voltage node and the ground. By using RRAM-CMOS -based inverters (2T2R primitive circuit), we are effectively capable of performing a memory read operation and having only temporary dynamic power dissipation (pMOS and nMOS conductive at the same time). The "memory read" operation for the case of 2T2R hybrid inverter is encoded into the inverter's threshold voltage, thus the input voltage point for which the inverter output is switching. Hence, not only we are reducing the power dissipation on the output part of the cell with smaller transitions of the matchline charging-discharging cycle, but also we are reducing the power dissipation of the ACAM cell during template matching operation. Although our proposed 9T4R cell requires more area per cell to implement arrays, in an assumed program-once-read-many application scheme (where the RRAM-based memory will be programmed once and then used for a long time before the next RRAM conductances update) the trade-off could be considered promising when we focus on the power consumption aspect of the ACAM for applications on the edge. Moreover, the use of RRAM-based ratio-based mapping of the analogue threshold values can enable the integration of a wider range of RRAM technologies, that showcase different ranges of viable resistive states and intrinsic behaviour, into the proposed 9T4R (compared to other ACAM pixel designs that base their memory read operation on voltage dividers) thus making the proposed design potentially more technology-agnostic. \par
For the comparison of our proposed 9T4R cell design with the current state-of-the-art, both cell types were designed and simulated in the same commercially available 180 nm technology and the same assumptions for the integrated RRAM technology were used towards adapting the MOSFET devices to the RRAM specifications (e.g. high current pass-through the accompanying MOSFET devices). The implementation employs 5 V components of the 180 nm library to enable the use of high electroforming voltage necessary for the RRAM technology under investigation. The precharge/enable signals period was set for $t_{clock}$ = 15 ns with $\frac{1}{3}\times t_{clock} = t_{evaluate}$ = 5 ns during assigned for memory read operation and $\frac{2}{3}\times t_{clock} = t_{initialise}$ = 10 ns) for pre-charging/initialisation (see Fig. \ref{fig:9T4R_Cell_Schematic}(b)). We are assuming that $t_{initialise}$ = 10 ns of precharge are useful in case a large ACAM matchline needs to be precharged to $V_{DD}$. The memory part of both cells is power gated and is turned off during the matchline initialisation phase to minimise the power consumption. \par 
\begin{table}[htbp]
\caption{9T4R Pixel Comparison with State-of-the-Art.}
\begin{center}
\begin{tabular}{|c|c|c|c|c|c|c|}
\hline
\textbf{Energy}&\multicolumn{6}{|c|}{\textbf{ACAM Pixel Designs}} \\
\cline{2-7} 
\textbf{Figures} & \multicolumn{2}{|c|}{\textbf{\textit{6T2R \cite{Li2020_HP_ACAM}}}}& \multicolumn{2}{|c|}{\textbf{\textit{3T1R} \cite{Agwa2023_3T1R_ACAM}}}& \multicolumn{2}{|c|}{\textbf{\textit{9T4R}}} \\
 \hline
 Input range & Low & High & Low & High& Low & High\\ 
 \hline
 Match & \multicolumn{2}{|c|}{2.25 pJ} & \multicolumn{2}{|c|}{35.5 fJ} & 152 fJ & 168 fJ \\ 
 \hline
 Mismatch & 479.2 fJ & 3.84 pJ & 11.39 fJ & 67 fJ & 30 fJ & 130 fJ \\   
 \hline
\multicolumn{7}{l}{}
\end{tabular}
\label{tab:table_pixel_energy_comparison}
\end{center}
\end{table}
As we can see from the figures in Table \ref{tab:table_pixel_energy_comparison}, for both match and mismatch, the proposed 9T4R exhibits better power consumption over the evaluation period (which was set for one $t_{clock}$ = 15 ns period). More specifically, for the case of match, 9T4R consumes on average only $E_{match}$ = 0.16 pJ while the 6T2R consumes 2.25 pJ. Similarly for the case of mismatch, 9T4R consumes on average only $E_{mismatch}$ = 0.08 pJ while the 6T2R state-of-art consumes on average 2.1 pJ (minimum energy approximately 0.479 pJ and maximum energy approximately 3.84 pJ). To better assess the trade-off of our proposed cell design, another comparison point with a recent ACAM cell is provided in Table \ref{tab:table_pixel_energy_comparison}. The cell is a 1T1R+2T ACAM cell proposed by \cite{Agwa2023_3T1R_ACAM} that is showcasing competitive energy results in both match and mismatch cases compared against the 6T2R and the proposed 9T4R. The trade-off of this higher energy efficiency is that the 1T1R+2T cell showcases a restricted capability of forming arbitrary matching windows compared to both 6T2R and 9T4R cells. The 1T1R+2T cell has been recreated to 180 nm and 5 V components and simulated with the same timing and input constraints for the purposes of this comparison. It is worth noting that especially since the memory read part is based on read operation on RRAM devices, the energy dissipation is strongly correlated with the memory state of the ACAM cell (the low and high bound mapped to RRAM states) and the input stimulus, that controls the voltage divider for the case of the 6T2R pixel and the hybrid inverter for the case of 9T4R (due to source degeneration of the MOSFET devices). Thus, it can be observed that a variability of energy dissipation exist based on the exact configuration of the cell's RRAM conductance and the input stimulus it receives. To overcome this issue and provide an indicative approximate energy figure, we are calculating the energy dissipation during a clock cycle (for our testbench this is $t_{clock}$ = 15 ns) and make this calculation for different configurations of the testbench. We are calculating an average (as shown in values of Table \ref{tab:table_pixel_energy_comparison}) based on our calculations which is indicative of each cell's performance. The matchline discharge/charge mechanism consumes approximately the same power between the different designs. The 6T2R consumes approximately 0 J in case of a match while the opposite occurs for the 9T4R cell, which consumes approximately 0 J in case of a mismatch. This is due to the event generation nature of the 9T4R comparison mechanism when we have a hit. The 9T4R cell's response takes the form of a short pulse that is "fired" when a match occurs instead of keeping the matchline precharged only to be discharged for the case of a mismatch. Thus, although the 9T4R is larger in area than the 6T2R, it showcases advantages in terms of energy efficiency. This efficiency of the 9T4R cell design can be further enhanced in system-level applications especially for in-sensor or near-sensor approaches for bio-signal classification, an area that is known to have sparse event while requires always-on sensing and classification of the input stimulus \cite{Rovere2018}. \par

\section{Process Variability Effects on 9T4R Cell} \label{sec:processVariability9T4R}
Since we are aiming at performing analogue memory-centric computing, we need to identify the main limitations of the proposed technology under the prism of process mismatch variability. Thus, we are performing a PVT corners analysis to observe the worst-case scenario performance of the proposed 9T4R cell and identify the techniques we can apply to mitigate the CMOS-related process variability effects. \par
\begin{figure}[t]
\includegraphics[width=\linewidth]{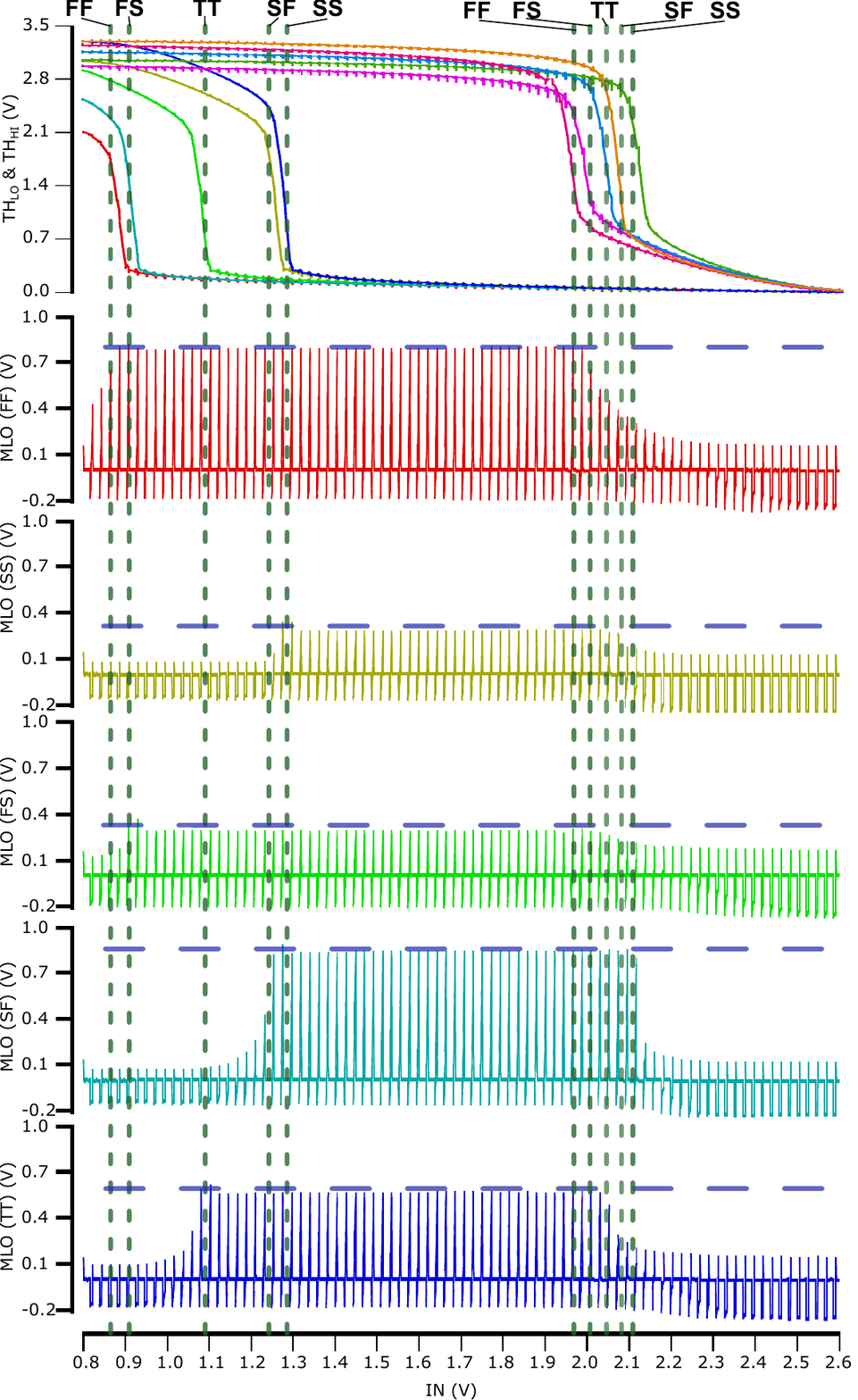}
\centering
\caption{In this figure, process variation is tested through corner analysis on the 180 nm technology. In the upper trace, the voltage threshold of the hybrid RRAM-CMOS inverters are shown for the five corner cases tested (FF, SS, FS, SF and TT). In the lower 5 traces, the output of the cell is shown for all corner cases. The results were captured from a transient simulation where an input ramp is used sweeping from $V_{IN}^{min}$ = 0.8 V to $V_{IN}^{max}$ = 2.6 V for the time duration shown. An enable signal is also used to emulate the match event generation operation ($V_{EN}^{max}$ = 3.3 V, $V_{EN}^{min}$ = 2.2 V and $t_{EN}$ = 12 ns). The input and control signals are similar to the simulation shown in Fig. \ref{fig:9T4R_matchin_window_response}. The results indicate that there is variability for the voltage threshold of the inverters that result in a matching window shift as well as 3T part resistivity variability that results in shifts of the match event output amplitude.}
\label{fig:process_variation_sim}
\end{figure}

The corner analysis is focused mainly on two aspects of the 9T4R, the degree of shifting of the voltage threshold of the two hybrid RRAM-CMOS inverters and the resistivity of the 3T comparison part in the circuit (as shown in Fig. \ref{fig:9T4R_Cell_Schematic}(d)). The shifting of the voltage threshold results in alteration to the matching window configuration while variability in the resistivity of the 3T part results in changes to the amplitude of the match event charging characteristics. The process variability effects can be seen in Fig. \ref{fig:process_variation_sim} for the cases of FF, SS, FS, SF, and TT with regards to the CMOS components used. The matching window configuration can be adjusted by further calibration of the RRAM weights while the mitigation of the increased/decreased resistivity of the 3T comparison part can be calibrated through appropriate configuration of the enable $EN$ signal's voltage levels and the enable signal's input evaluation period $t_{EN}$. \par 
In Fig. \ref{fig:process_variation_sim} simulation results from the corner analysis performed on the 9T4R cell are showcased. In the upper trace, the voltage threshold of the two hybrid inverter branches are shown while, in the lower 5 traces, the output of the cell (3T comparison part output circuit) are shown. As can be observed, for the different corners and the same RRAM configuration, the process variability affects both the matching window range and the amplitude of the match event output. The matching window shift effect can be mitigated by appropriately calibrating the RRAM devices. With regards to the output amplitude variability, this can be controlled through the enable signal. Calibration of all different variabilities in an array could be difficult to perform through a common enable connection, thus multiple connections (per row or per column) could be used to provide some tuneability to the array. Additionally, the use of an extra RRAM device could be used to calibrate the resistivity of the 3T part. \par
\section{ACAM System for Template Matching} \label{sec:prototypeIC}
\begin{figure*}[t]
\includegraphics[width=17cm]{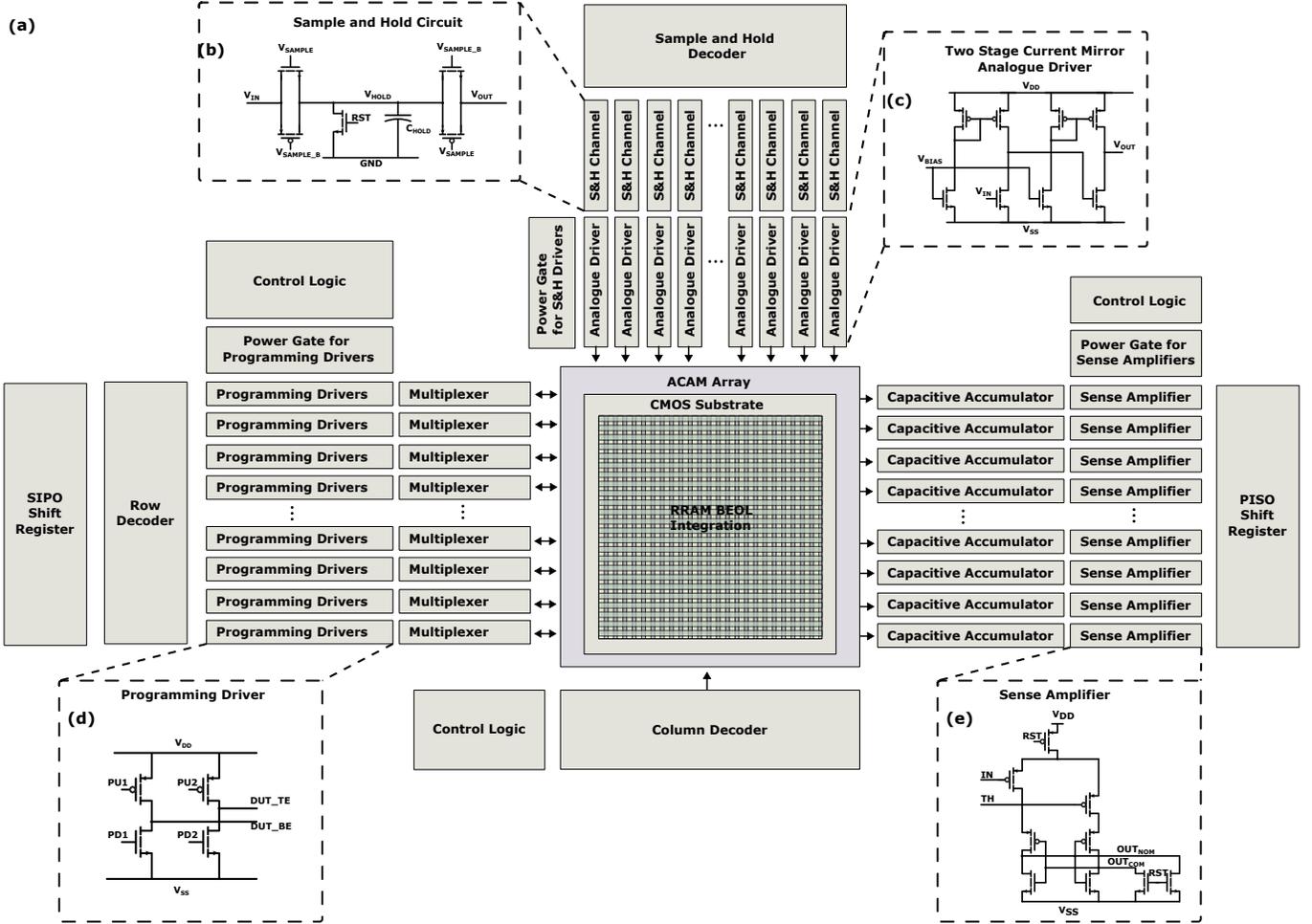}
\centering
\caption{(a) A system-level block diagram of the designed TXL-ACAM prototype IC is shown. At the centre of the block diagram is the main ACAM array while in periphery the circuits for capturing and driving the analogue input vector, supplying power and controlling the array as well as reading the ACAM are implemented. (b) Sample and hold circuit schematic. The S\&H converts the continuous analogue input into a real-valued query vector. (c) Analogue drivers for providing the query input to the array circuit schematic. (d) Programming circuit schematic. The programming circuits are responsible for providing the forming and programming voltages to the RRAM devices in the array. (e) Capacitor-based accumulator and Sense amplifier circuit schematic. The capacitive accumulator is adding all voltage packages generated by the match events per template. The total matchline charge accumulated is used as the input to the sense amplifier to be compared against a predefined threshold value above which a template match is triggered.}
\label{fig:Block_Diagram}
\end{figure*}

\begin{figure*}[t]
\includegraphics[width=15.7cm]{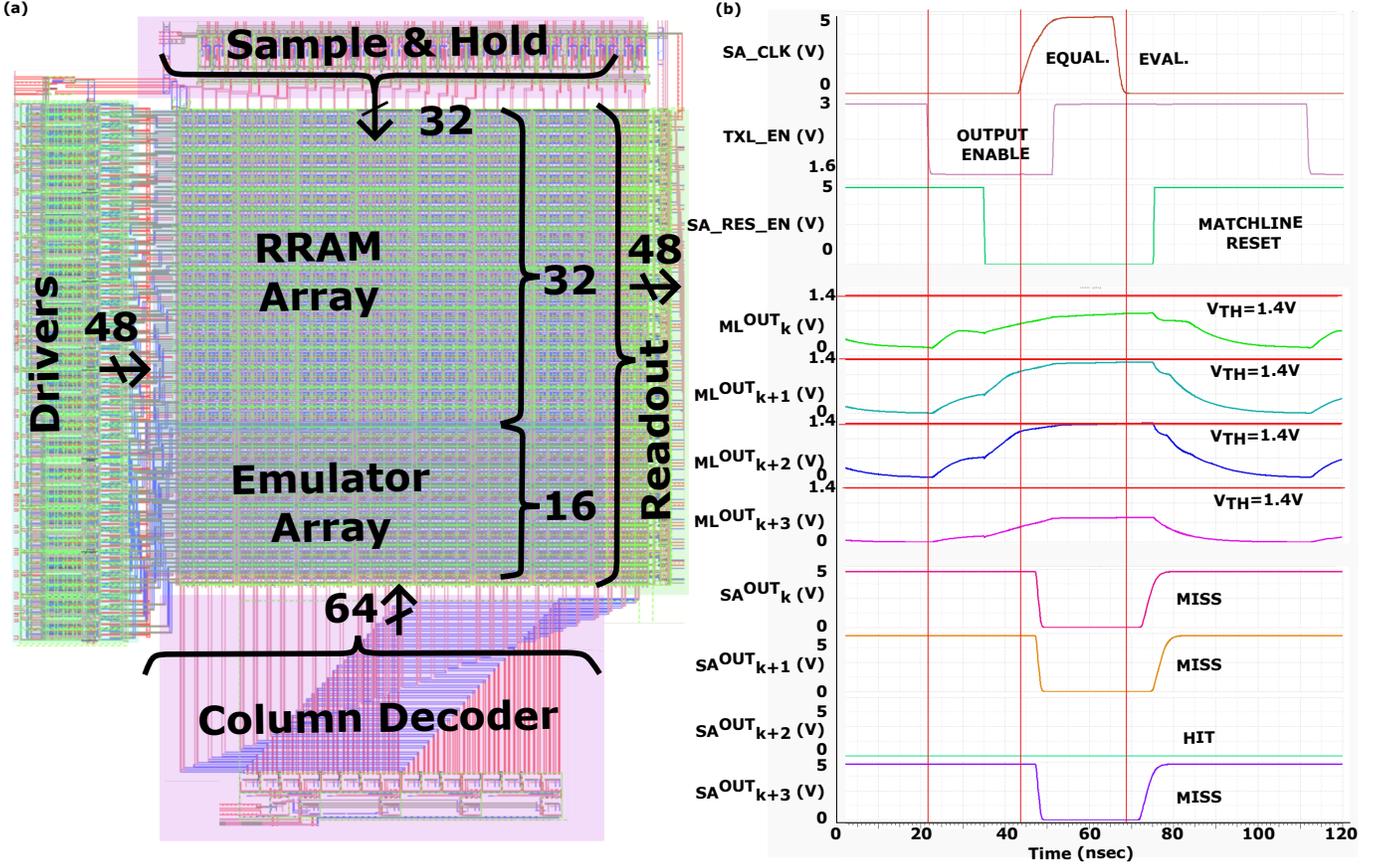}
\centering
\caption{(a) Layout design of the TXL-ACAM prototype IC for analogue template matching. The IC includes the main array closely integrated with sense amplifiers. From the top side, decoding circuitry with sample and hold array is used to sample an analogue input which is used as the analogue input vector for the 9T4R array. On the left side, a deserialiser (SIPO shift register) is used to load configuration bits and decoding logic is used to control the row drivers. The row drivers can access the RRAM devices of the array with either polarity (i.e. for SET and RESET conductance writing operations). The bottom side includes a column decoder with its routing network which is enabled when a writing operation is under way. The column decoder alongside the row drivers can select a standalone RRAM device during characterisation mode. On the left side, a serialiser (PISO shift register) is used to load the output of the sense amplifiers and shift all outputs through a single output pin. (b) Simulation results of the prototype IC implemented in 180 nm CMOS technology with the data-driven RRAM model used for testing based on our in-house technology \cite{Stathopoulos2017a, Messaris2017}. The $SA_CLK$ shows the sense amplifier clock signal controlling the equalisation-evaluation cycle of the comparator. The $TXL_EN$ is the array global enable signal that opens the output branch of cells enabling them to charge the matchline. The $TXL_EN$ has custom high and low level voltage values calibrated for fully breaking and partially making the output circuit. The $SA_RES_EN$ controls the leakage path that resets the matchline between subsequent pattern search operations. The signals $ML^{OUT}_{k}-ML^{OUT}_{k+3}$ represent a part of the 48 matchlines that for this example operation centre around the hit event. The signals $SA^{OUT}_{k}-SA^{OUT}_{k+3}$ are the sense amplifier outputs of the respective matchlines. For our design, the $SA^{OUT}=0 V$ level indicate a pattern hit while a $SA^{OUT}$=5 V level indicates a pattern miss. In this specific example showcased, it is worth noting that multiple initialisation and evaluation phase are being performed using the same input query values. Thus, sense amplifier that matches in the example operation also matched in the previous operation and this resulted in showcasing a continuous fully discharged state. (c) Area utilisation of each block in the prototype IC design. Approximately two thirds of the total area is covered by the 9T4R-based array. The rest one third of the area is covered by the implemented peripheral circuitry such as row circuitry block implementing SIPO shift register, decoding, programming drivers and multiplexer, the sense amplifier block includes an RC circuit for accumulating and resetting the matchline charge, output buffers and the PISO shift register and the S\&H block includes the MIM-based capacitor sampling channels and analogue buffers and decoding. The circuit coverage percentage showcase the relative size of a specific circuit compared to the total circuit area coverage of the prototype IC. The calculation of the total circuit area does not include the routing that interconnects the core circuits.}
\label{fig:Core_IC_Layout}
\end{figure*}

In this section, we are showcasing a prototype IC design developed using the proposed 9T4R ACAM cell as its building block. The IC is designed using 5 V components of the above-mentioned 180 nm technology and uses a full analogue design flow for this fully custom ACAM macro cell. A 9T4R-based 32x48 ACAM array alongside the necessary peripherals to access the RRAM devices for programming and reading as well as for performing the parallel template search and match operation was integrated into a single silicon IC with all CMOS circuits being implemented as computing substrate and the RRAM devices are added with our in-house Back-End-Of-Line (BEOL) process. A block diagram of the prototype IC is shown in Fig. \ref{fig:Block_Diagram}(a) while the physical layout of the implemented prototype IC is shown in Fig. \ref{fig:Core_IC_Layout}(a) alongside some notes with regards to IC's general floorplan. \par
Since the prototype IC is aimed at analogue template matching based on analogue information captured through a single input channel, a custom S\&H circuit to sample incoming continuous input is implemented. This is widely used for near-sensor edge computing such as spike-sorting in BMI, bio-signal classification applications, etc. Thus, the assumption followed for this IC design is that the front-end sensing and pre-processing circuitry should be supplying a pre-processed analogue signal to our proposed IC design. The 9T4R-based ACAM IC is capable of classifying the pre-processed input signal without requiring any analogue-to-digital conversion but by performing the massively parallel search and match operation of ACAM in the analogue domain. \par 
After the sampling operation of the continuous analogue voltage input is complete, custom analogue CMOS buffers are used to supply the input samples to the ACAM array. The captured analogue samples are organised as an input query vector which is compared with the stored analogue templates. Due to the nature of the proposed ACAM design that generates a matchline charge event per cell that has a match, a custom capacitor-based accumulator circuit for gathering the match event contributions is used per template. The accumulator circuit is implemented per row to add the charging effects of all cells per template. The accumulated charge per template represents a total similarity rate for each template compared to the provided input query vector. The accumulator circuit is connected to the sense amplifier and the total matchline charge voltage value is compared to the sense amplifier's threshold. The circuit consist of 1 capacitor that stores charge, 1 resistor that is enabled to reset the capacitor through leakage and 2 nMOS devices controlling the capacitor and resistor. The sizing of the components has been decided based on the operating point of the array, thus at $V_{DD}$ = 3.3 V and $F_{array_en}$ = 66 MHz. For this prototype TXL-ACAM IC, the leakage resistor and accumulator capacitor have been set to $R_{leakage} = 100 K\Omega$ and $C_{accumulator} = 100 fF$, respectively. In Fig. \ref{fig:Block_Diagram}(e) the circuit for the accumulator alongside the sense amplifier is showcased. The sense amplifier is used to digitise the ACAM read operation and provide the output of the ACAM array by applying a threshold matching operation. Every matchline that exceeds a specified level of matching, which is based on the number of match events generated for each matchline, is considered to have a matching template to the input query. The sense amplifier design is a voltage-mode dynamic latch comparator variant customised to sense the range of matchline voltage readout. The sense amplifier's threshold is arbitrarily selected based on the matchline charging dynamics. The matchline dynamics can be calibrated through the parasitic capacitance of the matchline in addition to matchline capacitor-based accumulator circuit, the voltage supply of the array (nominally set to 3.3 V) and the timing of the array global evaluation enable trigger signal. The sense amplifier threshold voltage level is calibrated during the TXL-ACAM programming phase where the array is programmed with the templates. The outputs of the sense amplifiers are connected to a custom Parallel In Serial Out (PISO) shift register which is loading all sense amplifier outputs in parallel and then serially shifting the outputs to a single pin. The PISO uses separate load and shift controls to easily configure the output readout operation. \par 
Similarly, with regards to the necessary configuration digital signals used to control the ACAM IC, a custom Serial In Parallel Out (SIPO) shift register. For each operation, the SIPO is loaded serially with all necessary control bits and then the connection between the SIPO and the rest of the ACAM IC is enabled to configure a specific state of operation. The connection between the SIPO and the rest of the array is controlled by tri-state buffers that can be configured to leave the connection floating during the SIPO loading operation. Custom drivers to enable programming and characterisation of the RRAM devices, alongside the control logic, are also included to provide the nominal 3.3 V supply to the array during the parallel search and match operation as well as programming voltages (up to 5 V) when a standalone RRAM is accessed through the per cell access transistors. A block diagram of the TXL 9T4R-based ACAM IC is shown in  Fig. \ref{fig:Block_Diagram}(a), where many of the main circuits used for the peripherals are also shown. The schematic of the sample and hold circuit, the analogue drivers, the programming circuit and the sense amplifiers are all shown in Fig. \ref{fig:Block_Diagram}(b),(c),(d),(e), respectively. \par
\begin{table*}[htbp]
\caption{TXL-ACAM IC Comparison with State-of-the-Art}
\begin{center}
\begin{tabular}{|c|c|c|c|c|c|c|c|}
\hline
\textbf{Prototype IC}&\multicolumn{7}{|c|}{\textbf{ACAM IC Designs}} \\
\cline{2-8} 
\textbf{Implementation} & \textbf{\textit{\cite{Li2020_HP_ACAM}}}& \textbf{This work} & 12T \cite{Koo_Match_Line_Clamping} & 3T1R \cite{Chang_3T_1R}& 2T2R \cite{Pan_2T_2R}& ASIC IMC \cite{Mingu_IM_RF}& DTIM \cite{Pedretti2021}\\
 \hline
 Type of CAM& Analogue & Analogue & Ternary & Ternary & Ternary& Binary & Analogue \\ 
 \hline
 CMOS & $22 nm$ & $180 nm$ & $28 nm$& $90 nm$& $65 nm$& $65 nm$& $65 nm$\\
 Technology  &   & (5 V  MOSFETs) & & & & & \\ 
 \hline
 Operating Voltage & $0.6 V$-$0.9 V$ & $3.3 V$& $1 V$& $1 V$& $0.7 V$& $0.75 V - 1 V$& $0.6 V$-$0.9 V$ \\ 
 \hline
 Memory & $TaO_{X}$-based & $TiO_{X}$-based  &  & & & & $TaO_{X}$-based \\
Technology & RRAM &  RRAM & CMOS & RRAM& RRAM& CMOS& RRAM \\
 \hline 
 IC & $12.5 um^{2}$ & $3.8 mm^{2}$  & N/A& N/A& N/A& $1.44 mm^{2}$& N/A \\ 
 Area    &   & (IO with core)  & & & & & \\
 \hline 
 Number of  & $1,032$ cells & $1,536$ cells  & $4,096$ cells& $4,096$ cells & $4,096$ cells & $131,072$ cells &$7,680$ cells \\
 ACAM Cells &  ($86\times12$ arr.) &  ($32\times48$ arr.) & ($64\times64$ arr.) & ($64\times64$ arr.)& ($64\times64$ arr.)& ($512\times256$ arr.)&($16\times480$ arr.) \\
 \hline 
 IC & $0.57 fJ$ & $185 fJ$ & $1.62 fJ$ & $0.51 fJ$ & $0.23 fJ$ & $150 fJ$ &$160 fJ$ \\ 
 Energy & (per search  & (per search & (per search& (per search& (per search&(per search&(per search \\ 
& per cell)  & per cell) & per cell)& per cell)& per cell)& per cell)&per cell) \\ 
 \hline 
\multicolumn{7}{l}{}
\end{tabular}
\label{tab:table_ic_comparison}
\end{center}
\end{table*}
Towards simplifying the system's operations we implemented the IC capable of switching between two modes, the analogue template matching mode, where analogue input vectors is applied in parallel to all entries in the ACAM array and the results of this memory read are sensed by the sense amplifiers and serialised with the PISO output shift register, and the characterisation/programming mode, where the programming drivers are used to characterise/program individual RRAM devices. For the programming mode, row and column decoders are used to isolate specific RRAM devices for programming. \par 
The physical layout for the proposed 9T4R-based ACAM IC is shown in Fig. \ref{fig:Core_IC_Layout}(a), alongside some notation to provide more information on the sub-systems employed for the proof-of-concept design and showcase the approximate floorplan of the design. The physical layout of the proposed ACAM array was designed using a commercially available 180 nm technology for CMOS and our in-house RRAM technology based on Pt/$AlO_{x-2}$/$TiO_{x}$/Pt metal oxide bi-layer metal-insulator-metal (MIM) RRAM technology. The RRAM technology employed has been shown to exhibit strong multi-bit memory operation \cite{Stathopoulos2017a}. We are replacing some rows of the RRAM-based ACAM array with rows that include conventional polysilicon-based resistor devices (based on components from the same 180nm technology). This array configuration is implemented alongside the RRAM-CMOS array for calibration purposes and readout assist purposes. Thus, the emulator array can be used asreference point when reading a standalone RRAM device for characterisation purposes after the BEOL integration. Additionally, it is used as a testing array towards helping with characterising the TXL-ACAM behaviour when different RRAM technologies are integrated and tested. Thus, the total $48\times32$ ACAM array of this prototype IC has $32\times32$ RRAM-based rows and $16\times32$ polysilicon-based rows (the two sub-arrays are unified into a single array as shown in Fig. \ref{fig:Core_IC_Layout}(a)). As shown in Fig. \ref{fig:Core_IC_Layout}(c), the area utilisation of the main array covers the majority of the total IC area (approximately 66.3\% of the total area). From the implemented peripherals, the programming drivers and the S\&H channels are two of the largest block of the prototype IC with area usage of approximately 8.8\% and 6.3\%. It is worth noting that in Fig. \ref{fig:Core_IC_Layout}(c) the area coverage of each circuit compared to the total circuit area is assessed without including the routing area. Thus, the coverage percentage indicate the relative area utilisation of each type of circuit compared to sum of circuit area utilisation in the prototype IC. \par
At the top part of the IC, the input capture peripherals (part of the analogue front-end) are included. A decoder provides phase control for the sampling operation of the parallel S\&H circuit array. Based on the one-hot encoding generated by the digital control circuitry, the piecewise sampling of the analogue input is possible. Each sample is stored in one capacitor device (mimcap device). Transmission gates (TGs) are used to control the sampling of the S\&H circuits. The TGs are appropriately sized (W/L = $10 \mu m/600 nm$ for both pMOS and nMOS devices of the TG) to enabled the fast charging of the hold capacitor to appropriate levels of charge. When all different samples of the analogue signal have been acquired, then a parallel analogue buffing supports the transmission of the analogue vector into the array (for the pattern matching operation). For the case of the programming operation of the array, the input signal is set to logic high (for $V_{DD-TXL}$=3.3 V). The output of the array (when in normal pattern matching operation mode) is captured by a sense amplifier array. Each sense amplifier is designed as a dynamic latch comparator type of circuit effectively comparing the matchline voltage to a specific predetermined biasing voltage. Each sense amplifier outputs the results of this comparison. All sense amplifier outputs are connected in parallel to a serialiser circuit (effectively a Parallel Input Serial Output Shift Register). Thus, when the pattern recognition phase is completed, all the outputs of the sense amplifier array are loaded into the serialiser and then a bit-stream is sent through a single output pin by shifting the contents of the serialiser to this output pin. \par
In the characterisation mode (also referred to as programming mode), the TXL cell “disconnects” the RRAM devices from the rest of the cell and each RRAM device can be uniquely addressed through the use of accompanying “programming” nMOS devices. Effectively, during the programming/characterisation mode, the array becomes a 1T1R type of crossbar array, thus every TXL (9T4R cell) circuit becomes 2 1T1R cells. The access (for programming purposes) to each RRAM device is controlled by a 64-bit decoder which is physically placed at the bottom of the TXL array and not showcased as part of the normal programming circuitry (physically placed at the left of the TXL array). The programming logic used to generate the appropriate programming control signals and thus control the programming drivers and program/read multiplexing are implemented using (mostly) minimum size high threshold 5V MOSFET components (thus, W/L=$220 nm/600 nm$). Buffers are necessary to increase the driving strength between the minimum sized logic and the driver circuits, to effectively control the programming MOSFET connections to TE and BE (with regards to the RRAM devices under test). The size of the programming drivers depends on the 9T4R array size (and the resistive load that is expected to be encountered in the programming paths, especially for the ones at the opposite end of the array compared to where the programming circuits exist). The estimated size is W/L=$20 \mu m/600 nm$ for pMOS and W/L=$10 \mu m/600 nm$ for nMOS devices that comprise the programming drivers. \par
For design validation, we performed post-layout simulation on this proof-of-concept IC after extracting the resistive and capacitive parasitics of the physical layout. Since the RRAM model used to perform pre-layout simulations is not yet adapted for parasitics extraction, for the purpose of validating the main functionality of the design post-layout, we are replacing the RRAM components with polysilicon-based resistors in a similar manner with the polysilicon-based part of the initial IC which is used for calibration and RRAM read reference. Although this is not enabling us to include the full intrinsic characteristics of the RRAM devices into our design, we are using this method to provide a baseline system analysis to prove the functionality of our array design alongside the implemented peripheral and how they are affected by resistive and capacitive parasitics. An example operation of the analogue pattern matching operation is shown in Fig. \ref{fig:Core_IC_Layout}(b). More specifically, a part of the matchline readout vector alongside the sense amplifier classification and the array and sense amplifier control sequence is showcased. In this example operation, we are showcasing the main control signals required to read the TXL-ACAM array. The $SA_CLK$ control the equalisation-evaluation cycle of the sense amplifier, with equalisation setting up the sense amp into a low power unstable state and the evaluation enabling the sense amp to settle quickly depending on the voltage threshold and the matchline charge. The $TXL_EN$ is the global array enable signal that dictates the reading phase of the ACAM array during the pattern recognition phase. When the query is supplied to the array, $TXL_EN$ enables the charging of the matchline for each cell that has an input that falls between the match window, as determined by the RRAM values, by controlling the $M_{MEN}$ pMOS as shown in Fig. \ref{fig:9T4R_Cell_Schematic}(d). The $SA_RES_EN$ is a global matchline control signal that dictates the reset of the matchlines by discharging them through a specific rate, which is determined by a leakage resistor implemented per matchline. The output response of the TXL-ACAM array is shown in Fig. \ref{fig:Core_IC_Layout}(b). The matchline outputs $ML^{OUT}$ (for this example the subset of $ML^{OUT}_{k}-ML^{OUT}_{k+3}$ is shown) illustrate the charging-discharging cycle of the pattern search operation. The sense amplifier output $SA^{OUT}$ (for this example the subset of $SA^{OUT}_{k}-SA^{OUT}_{k+3}$ is shown) showcase the output of the dynamic latch-based circuit which effectively performs a comparison between the charging level of each matchline during the specific time when the sense amplifier evaluation starts against a provided threshold voltage $V_{TH}$ which is depicted alongside the $ML_{OUT}$ traces. For the case of our example operation, we are setting $V_{TH}=1.4V$. For our design, the $SA^{OUT}=0V$ level indicate a pattern hit while a $SA^{OUT}$=5 V level indicates a pattern miss. It is worth mentioning that for the specific part of the simulations showed, the same input query is used for multiple equalise-evaluate cycles of the sense amplifier. Thus, the matching line is showing only $0V$ since both the equalisation the evaluations are setting it to $0V$.\par
In Table \ref{tab:table_ic_comparison}, we are showcasing some indicative traits of the proposed TXL prototype IC compared to other state-of-the-art ACAM designs. We can see that our prototype IC is consuming more energy per classification per ACAM cell but it is worth noting that our implementation is using 180 nm CMOS technology and 5 V MOSFET devices to accommodate the BEOL RRAM technology employed for our design, while the objective of this prototype IC is to achieve high versatility and fine-tuning capabilities from the ACAM array. The nominal operation voltage of the peripheral circuitry is 5 V while the main array operates at 3.3 V. The design is aimed to be tested through an analogue stable technology before being transferred into a modern node with lower voltage supply. \par

\section{Conclusions \& Discussion} \label{sec:conclusions}
In this work, we propose a RRAM-based TXL-ACAM cell design for analogue template matching at the edge. This design is using a 9T4R cell configuration to enable energy efficient ACAM applications. The cell was designed and tested using a commercially available 180 nm CMOS technology and a RRAM model adapted on the physical characteristics of our in-house RRAM technology \cite{Messaris2017}. The proposed 9T4R ACAM cell design improves the tuneability of the matching windows in comparison to the state-of-the-art by using one hybrid RRAM-CMOS inverter per threshold. Also, it enables the easy accessing of each standalone RRAM device for electroforming and programming operations through the additional nMOS access devices. Finally, enables the configurable contribution of the cell to the charging matchline through the additional enable pMOS device at the comparator part of the cell. Although this results in a relatively large ACAM cell, it enables energy efficient operation due to its CMOS-oriented design of the RRAM-CMOS inverters that are used to map the matching window per cell. This results in an ACAM design that is a promising candidate for extreme edge computing. \par
Furthermore, a prototype IC design aimed at testing the novel 9T4R pixel cell in an ACAM macro implementation, co-designed with custom peripherals necessary for performing analogue template matching and RRAM device programming, is designed and tested in simulation using a commercial 180 nm technology. The IC was designed using 5 V MOSFET components from the 180 nm library and was fully custom-made following the analogue design flow in Cadence Virtuoso environment. The IC was tested with RC parasitic extracted components through post-layout simulations towards further validating the proper circuit behaviour of each IC block and estimating the performance of the TXL-ACAM. \par
The findings in this work illustrate that the proposed 9T4R-based TXL-ACAM could be considered a promising candidate for integration into near-sensor systems for analogue template matching at the edge due to its energy efficiency and enhanced tuneability capabilities. Such solutions could greatly benefit bio-signal processing either near-sensor or in-sensor using energy efficient memory-centric accelerators and template matching engines. The design and implementation of the showcased prototype IC using high threshold 5 V components enables the easy integration of different RRAM devices. Additionally, further design effort to minimise and optimise the peripherals will achieve a more compact prototype IC design. Although currently the full 5 V design was selected as a versatile and extra safe method for integration with a multitude of RRAM devices, the transfer to 1.8 V components will also be beneficial for power and area. Furthermore, transferring the design to a smaller MOSFET technology node, such as 90 nm or 65 nm, will further help with power and area. Through the use of smaller CMOS technology nodes we could have lower voltage and IC area required for the same array size, assuming that the MOSFET devices impose the array density limit. To apply a further miniaturised CMOS node update, RRAM technology needs to be sufficiently operated at the lower operating voltages of said CMOS nodes. Thus, further exploration of RRAM device technologies, in the form of different device's stack configurations, and its integration with CMOS will drive the miniaturisation of the current prototype IC. Furthermore, it is useful to consider that larger nodes such as 180 nm enable the design of more stable analogue circuit that can be used for classifying partial match per cell in ACAM systems. Additionally, higher static leakage could require additional calibration for the matchline charging but it could also be mitigated for the 2T2R branches due to higher resistance in the $V_{DD}$-$GND$ path. Finally, further investigation of how to integrate TXL-ACAM with additional components to test its architectural scalability for real-world applications will be important for better assessing the competitiveness of the design with the state-of-the-art. The test of scaled-up TXL-ACAM systems with real benchmarking, such as MNIST and CIFAR10, could provide an early estimation on the expected performance of the proposed design.  \par  
\bibliography{References_v2024_08_28_e4aGP.bib, Chang_3T_1R.bib, Koo_Match_Line_Clamping.bib, Mingu_IM_RF.bib, Montoye_2T_2R.bib, Pan_2T_2R.bib, Sachin_pcell.bib}
\begin{IEEEbiography}
[{\includegraphics[width=1in,height=1.25in,clip,keepaspectratio]{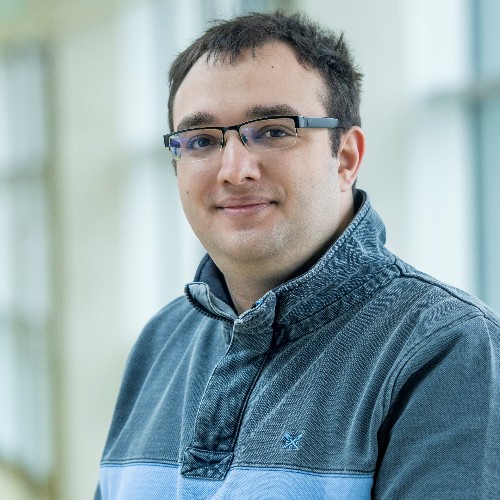}}]{Georgios Papandroulidakis}
is a Research Associate at the Centre for Electronics Frontiers (CEF), The University of Edinburgh (UK). He received his Diploma-MEng (in 2015) and MSc (in 2017) degrees on Electronics and Computer Engineering from Democritus University of Thrace, Greece, with his diploma and MSc thesis involving the design and simulation of memristor-based circuits and systems. He completed his PhD on reconfigurable RRAM-based circuit, systems and computer architectures in 2021 (University of Southampton, UK) working on hybrid RRAM-CMOS Threshold Logic Gates -based circuit and fast prototyping of hardware test setups for on-wafer measurements. He joined the CEF group as a Research Associate in 2021 working on areas of emerging electronics and their applications such as RRAM-based Circuit and Systems, In-Memory Computing, Approximate Computing, In-Sensor and Near-Sensor Pattern Classification, and Memory-Centric Hardware Accelerator Architecture Design. His most recent research focus is on ACAM neuro-inspired systems comprised of hybrid RRAM-CMOS circuit for energy efficient analogue pattern matching.
\end{IEEEbiography}
\begin{IEEEbiography}
[{\includegraphics[width=1in,height=1.25in,clip,keepaspectratio]{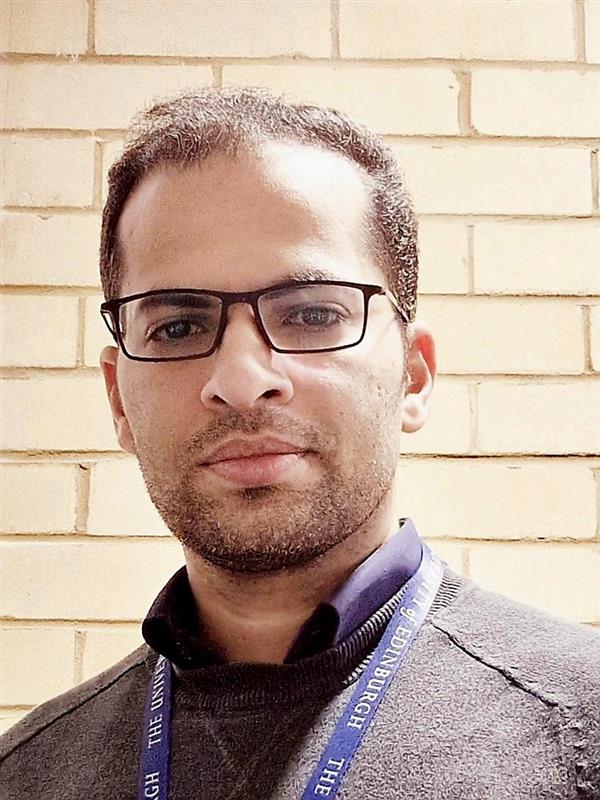}}]{Shady Agwa}
(Member, IEEE) is a Research Fellow at the Centre for Electronics Frontiers (CEF), The University of Edinburgh (UK). He received his BSc and MSc degree from Assiut University (Egypt), both in Electrical Engineering. He got his PhD in Electronics Engineering from The American University in Cairo (Egypt) in 2018. Following his PhD, he joined the Computer Systems Laboratory at Cornell University (USA) as a Postdoctoral Associate for two years. In 2021, Shady joined the Centre for Electronics Frontiers at the University of Southampton (UK) as a Senior Research Fellow and then as a Research Fellow at the University of Edinburgh (UK). His research interests span across VLSI and Computer Architecture for AI using conventional and emerging technologies. His work focuses on ASIC-Driven AI Architectures with extensive expertise in In-Memory Computing, Stochastic Computing, Systolic Arrays, Beyond Von Neumann Architectures, Memories and Energy-Efficient Digital ASIC Design.
\end{IEEEbiography}
\begin{IEEEbiography}
[{\includegraphics[width=1in,height=1.25in,clip,keepaspectratio]{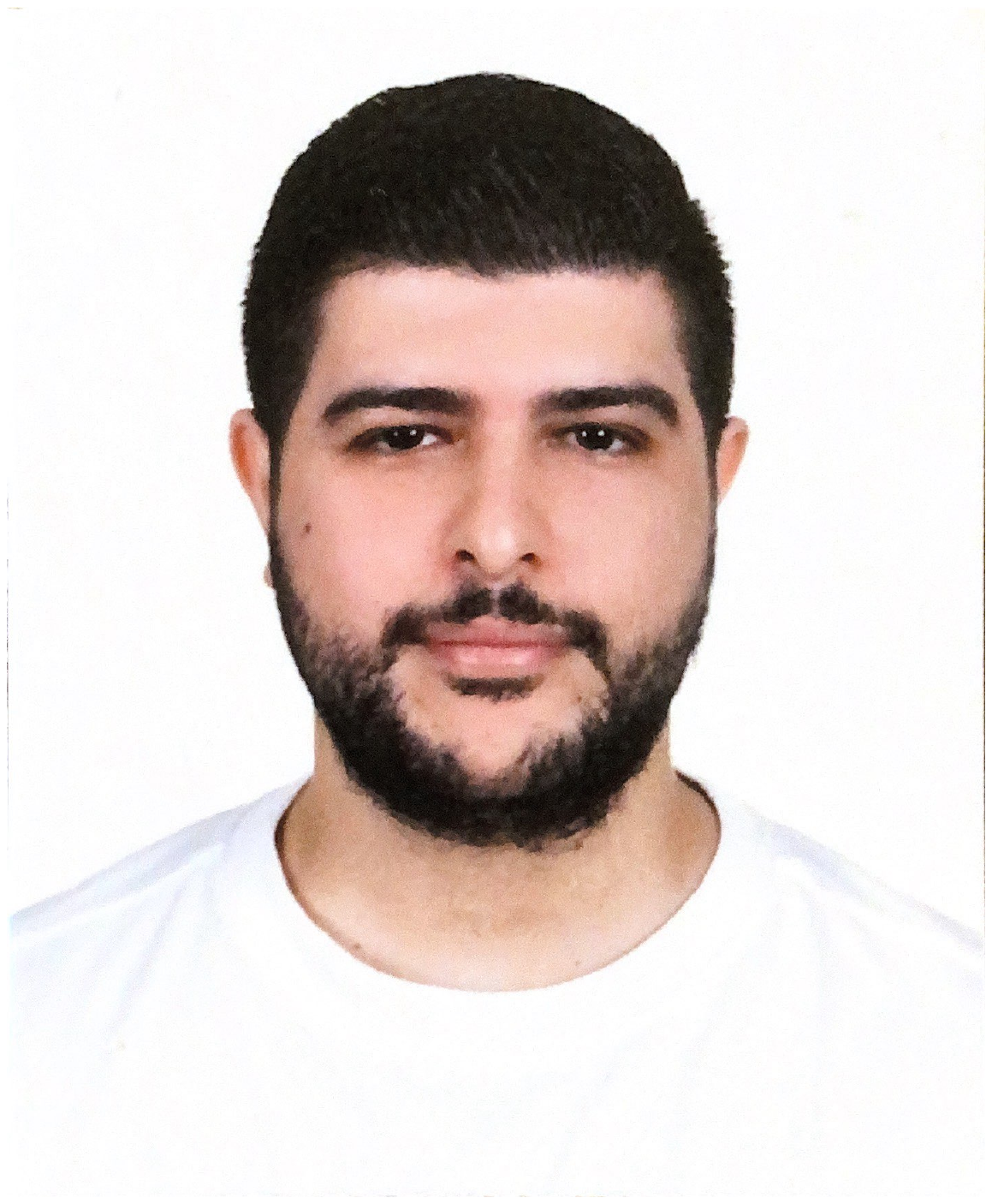}}]{Ahmet Cirakoglu} (Member, IEEE) is a PhD student at the Centre of Electronic Frontiers, Institute for Integrated Micro and Nano Systems, University of Edinburgh, United Kingdom. Prior to that, Ahmet was an RFIC design engineer at Phasor Solutions (now Hanwha Phasor) between 2017 and 2020. He received his M.Eng degree in Electrical and Electronic Engineering from Imperial College London in 2017 and M.Sc degree in Machine Intelligence for Nano-electronic Devices and Systems from University of Southampton in 2022. His research interests include design of radiation hardened, mixed-signal AI hardware accelerators, memory design with emerging non-volatile RRAM technology, electrical characterization and modelling or RRAM devices, in-memory computing and analogue circuit design.
\end{IEEEbiography}
\begin{IEEEbiography}
[{\includegraphics[width=1in,height=1.25in,clip,keepaspectratio]{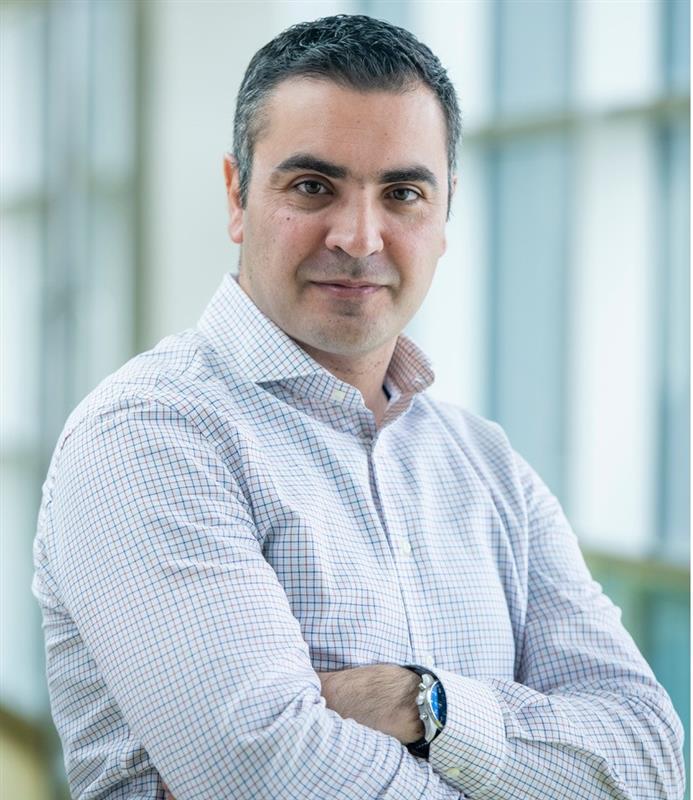}}]{Themis Prodromakis}
(Senior Member, IEEE) received the bachelor’s degree in electrical and electronic engineering from the University of Lincoln, U.K., the M.Sc. degree in microelectronics and telecommunications from the University of Liverpool, U.K., and the Ph.D. degree in electrical and electronic engineering from Imperial College London, U.K. He then held a Corrigan Fellowship in nanoscale technology and science with the Centre for Bio-Inspired Technology, Imperial College London, and a Lindemann Trust Visiting Fellowship with the Department of Electrical Engineering and Computer Sciences, University of California at Berkeley, USA. He was a Professor of nanotechnology at the University of Southampton, U.K. He holds the Regius Chair of Engineering at the University of Edinburgh and is Director of the Centre for Electronics Frontiers. He is currently a Royal Academy of Engineering Chair in emerging technologies and a Royal Society Industry Fellowship. His background is in electron devices and nanofabrication techniques. His current research interests include memristive technologies for advanced computing architectures and biomedical applications. He is a fellow of the Royal Society of Chemistry, the British Computer Society, the IET, and the Institute of Physics.
\end{IEEEbiography}

\end{document}